\documentstyle[aps,epsfig]{revtex}
\newcommand {\nb}{\nonumber\\}

\newcommand {\trimer}{$^4${He}$_3$ }
\newcommand {\trimerecc}{$^4${He}$_3^*$ }
\newcommand {\dimer}{$^4${He}$_2$ }

\newcommand {\be}{\begin{equation}}
\newcommand {\ee}{\end{equation}}
\newcommand {\bea}{\begin{eqnarray}}
\newcommand {\ea}{\end{eqnarray*}}
\newcommand {\ba}{\begin{eqnarray*}}
\newcommand {\eea}{\end{eqnarray}}
\newcommand {\bra}{\langle}
\newcommand {\ket}{\rangle} 
\newcommand {\rv}{{\bf r}}

\newcommand {\yv}{{\bf y}}
\newcommand {\xv}{{\bf x}}

\newcommand {\ab} {{\it{ab initio}}}
\newcommand {\au} {a.u.}

\title{Variational description of the Helium trimer using 
correlated hyperspherical harmonic basis functions}

\author{P. Barletta}
\address{Department of Physics and Astronomy, University College London, London WC1E 6BT, UK}
\author{A. Kievsky}
\address{Istituto Nazionale di Fisica Nucleare, Piazza Torricelli 2, 56100 Pisa, Italy}

\begin{document}

\maketitle


\begin{abstract}
A variational wave function constructed with correlated Hyperspherical
Harmonic functions is used to describe the Helium trimer. This system
is known to have a deep bound state. In addition, different potential
models predict the existence of a shallow excited state that has
been identified as an Efimov state. Using the Rayleigh-Ritz variational
principle the energies and wave functions of both bound states have been
obtained by solving a generalized eigenvalue problem. The
introduction of a suitable correlation factor reduces considerably
the dimension of the basis needed to accurately describe the 
structure of the system. The most recent helium-helium interactions have been
investigated.

\end{abstract}

\section{Introduction}
\medskip

The system He-He is greatly interesting both from a theoretical and an 
experimental point of view, and it has been object of intense
investigations in the last years. Despite its simplicity, 
it is not easy to determine whether it supports or not a bound state. 
Experimentally, usual spectroscopy techniques are not suitable to its study, 
and only recently~\cite{luo2,schol,schol2,gris} diffraction experiments 
proved its existence, with a direct measurement of its bond length $<R>$.
Its binding energy has been estimated through the relation
$|E_b|\approx\hbar^2/4m<R>^2$ and the $s$-wave scattering length has been
estimated as a$_0 \approx 2<R>$. The most recent values for these quantities
have been quoted in Ref.~\cite{gris}
after a new determination of the bond length by diffraction from a
transmission grating. They
are $<R>=98\pm 8$ \au, $|E_b|=1.1+0.3/-0.2$ mK and 
a$_0=197+15/-34$ \au.

Theoretically difficulties in the description of the \dimer
arise because the He-He interaction results from 
the subtraction of the huge energies of the separated atoms, which are only 
slightly different. Moreover, a $1\%$ decrease of the strength of the 
interaction makes the system unbound. 
As a result of the continuous refinement in the past years of both experimental
data and electronic structure computational techniques, several potential 
curves for He-He appeared in literature. Most of them are presented and 
compared in an article review by Janzen and Aziz  ~\cite{aziz5}. However,
 due to the vivid interest in \dimer, and to the difficulties in making a
 really accurate \ab \ potential, newer and more accurate potentials 
have recently appeared in literature. 
Among the ones described in Ref.~\cite{aziz5}, the potentials called 
HFDB \cite{aziz3}, LM2M2 \cite{aziz1}, TTY \cite{tang1} 
have been widely used in helium 
cluster calculations. Furthermore, two more up-to-date curves, 
namely SAPT1 and SAPT2 \cite{aziz4} are now available. 
The latter is believed to be the 
most accurate characterization of the He-He interaction yet proposed. These 
potentials are constructed on a completely \ab \ calculations made by Korona 
et al.~\cite{kor1}, using infinite order perturbation theory (SAPT) and very 
large orbital basis set. In addition, retarded dipole-dipole dispersion 
interaction is included over the range $0-100000$ \au (SAPT1), or the more 
appropriate 10-100000 \au (SAPT2). 

All these five interactions
support only a single bound state of zero total angular momentum.  
In Table I we summarize the different characteristics of each 
potential, as well as salient properties of the associated bound state. 
The latter has been computed solving the two-body Schr\"odinger
 equation by means of the Numerov algorithm . We used the value 
$ \hbar^2/m = 43.281307$ K \au$^2$.
>From the table we can immediately 
see that the five potentials do not differ qualitatively among each other,
though there is a spreading in the binding energy of the dimer $|E_b|$ of
$0.51$ mK. The SAPT potentials predict the highest binding energies whereas
the LM2M2 and TTY predictions are very close to each other and are the lowest 
ones. The differences observed in the binding energy are reflected in the mean 
values of the radius, $<R>$ and $<R^2>$, as well as in the scattering length
 a$_0$. 
The estimate bond length $<R>$ can be directly compared 
to the experiment and can be also used for an estimation of the binding 
energy $|E_b|$ and the scattering length a$_0$ through the relations given in
the first paragraph. Those values are shown in the last two rows of
Table I and reasonably agree with the estimation of 
Ref.~\cite{gris}, in particular the results obtained with the
LM2M2 and TTY potential are inside the quoted errors. 
We can also observe that the system is strongly 
correlated as its binding energy $|E_b|$ results from a large cancellation 
between the kinetic $<T>$ and potential energy $<V>$. Its spatial 
extension is considerably bigger than the range of the potential, as it is
shown in Fig.1 where the LM2M2 potential and the dimer bound state wave 
function $\Phi_d$ are plotted.  Finally, the scattering length a$_0$ 
of the system is bigger than the range of the potential 
by an order of magnitude. All these features characterize the \dimer as 
the weakest, as well as the biggest, diatomic molecule found in nature so far. 
Moreover the bound state at practically zero energy suggests the possibility 
of observing an Efimov-like state in the triatomic compound ~\cite{ef2,ef1}.

Along with the observations of small clusters of helium atoms, different
theoretical methods have been used to study the properties of such
systems. From the beginning it has been clear that standard techniques 
could have problems to describe those highly correlated structures and,
accordingly, more sophisticated methods have been applied. In 
Ref.~\cite{lew} the diffusion Monte Carlo (DMC) method was used to
describe the ground state of He molecules up to 10 atoms. The \trimer
has been extensively studied by different methods (see Ref.~\cite{sof}
and references therein). Theoretically it has been shown that the trimer
has a deep ground state of about $126$ mK and a single $L=0$ 
excited state of about $2$ mK. There are not bound states with
$L>0$  ~\cite{bruch}. In Refs.~\cite{esry,gnt2} the $L=0$ excited state has 
been studied, in particular looking at those characteristics 
identifying an Efimov state.
In fact this state has the property of disappearing when the 
 interaction strength is tuned with a parameter $\lambda$. 
For example, the excited state exists in the very narrow interval
$0.97 \lesssim \lambda \lesssim 1.2$ for the LM2M2 potential. 
Though with slightly different values of $\lambda$, the 
same property holds for the other potentials mentioned above. Therefore 
the helium trimer gives the unique possibility of observing the Efimov effect,
as the narrow range in $\lambda$ where the excited state \trimerecc appears
contains the physical case $\lambda=1$.

In the present paper a set of correlated basis functions is used to
describe the \trimer molecule. The correlated hyperspherical harmonic (CHH) 
basis has been applied successfully in the ground state description
of light nuclei~\cite{chh}. Similarly to the cluster of helium, these systems
 are strongly correlated due to the high repulsion of the nucleon-nucleon 
potential at short distances. Essentially the method consists in a 
decomposition of the wave function in terms of the hyperspherical harmonic 
(HH) basis multiplied by a suitable correlation factor which takes into 
account the fact that the probability of finding any pair of atoms at
distances smaller than $3$\ \au\   is practically zero. The correlation
factor has been taken as product of one dimensional correlation
functions $f(r)$ (Jastrow type). In Fig.1 $f(r)$ is compared to the dimer
wave function, showing that both have the same short range behaviour.

The variational description of the trimer using the CHH
basis is twofold. Firstly we would like to evaluate the
capability of the correlated basis functions to describe a strongly
correlated system.
Special attention will be given to the convergence pattern of the
energy for both the ground and excited states. 
In Ref.~\cite{sof} calculated  binding
energies of the ground and excited states of the trimer obtained by
different groups are given in correspondence with different interactions.
The solutions of the Faddeev equations as well as variational methods
and adiabatic approaches have been used in those calculations. For
the very shallow excited state of the trimer, only few results using the
 variational method have been reported so far, showing the difficulty of
 describing this state with the required accuracy using such a technique.
In the present work we will show that it is possible to obtain high 
precision upper bound estimates and wave functions for both the ground and
 excited  states by solving a generalized eigenvalue problem. Moreover,
a detailed study of the wave function will be performed. In particular 
the tail of the wave function will be analyzed with the extraction of the
asymptotic constants. The second motivation of the present work 
regards the extension of the method to describe larger systems. In fact
a complete study of the ground state and excited states of the 
tetramer has still to be performed. In this context the variational
technique is promising and the present study should be considered
a first step along this direction.

The paper is organized as follows. In the next section a discussion
of the CHH basis for the systems of three atoms is given. The
numerical results for the binding energy of the two bound states
are given in Sect. 3. Some properties of the wave functions and the
asymptotic constants are calculated in Sect. 4 whereas the main
conclusions as well as some perspectives for the extension to
larger systems are given in the last section.

\section{CHH basis}

In the present study the interaction between three helium atoms is taken
as a sum of three pairwise potentials. The Hamiltonian of the
system will be
\be
H=T+\sum_{i<j}V(i,j)
\label{ham}
\ee
where $T$ is the kinetic energy operator and $V(i,j)$ is the He-He
interaction that in the present work will be taken as one of the
potentials mentioned in the previous section.

Considering the helium atom as a spinless boson, the wave function
for three identical spinless bosons can be written as a sum of
three Faddeev--like amplitudes
\be
\Psi = \psi(\xv_1,\yv_1) + \psi(\xv_2,\yv_2) 
+ \psi(\xv_3,\yv_3) 
\label{ampl}
\ee
where the sets of Jacobi coordinates ($\xv_i,\yv_i$) ($i,j,k=1,2,3$ cyclic)
are:
\bea
\xv_i & = & {1\over \sqrt{2}} (\rv_j - \rv_k)  \nb
\yv_i & = & {1\over \sqrt{6}} (\rv_j + \rv_k - 2 \rv_i)
\label{jacs}
\eea
Each $i$--amplitude has total angular momentum $LM$ and can be 
decomposed into channels
\be
\psi(\xv_i,\yv_i)=\sum_\alpha \Phi_\alpha(x_i,y_i)
[Y_{\ell_\alpha}({\hat x}_i)Y_{L_\alpha}({\hat y}_i)]_{LM}
\label{twod}
\ee
A symmetric wave function requires $\ell_\alpha$ to
be even. Moreover $\ell_\alpha+L_\alpha$ should be even for positive
parity states. 

Let us introduced the hyperspherical variables
\be
  x_i=\rho\cos\phi_i, \quad y_i=\rho\sin\phi_i
\ee
where $\rho$ is the hyperradius which is symmetric under any permutation of 
the three particles and $\phi_i$ is the hyperangle. In terms
of the interparticle distances $r_{ij}=|\rv_i-\rv_j|$ the hyperradius reads:
\be
  \rho={1\over \sqrt{3}}\sqrt{r^2_{12}+r^2_{23}+r^2_{31}} \ .
\ee
Using the set of coordinates 
$[\rho,\Omega_i]\equiv[\rho,\phi_i,\hat{x}_i,\hat{y}_i]$, the volume element 
is $dV={\rho}^5 d \rho d\Omega_i ={\rho}^5 d \rho \sin^2{\phi_i} 
\cos^2{\phi_i} d\phi_i  d\hat{x}_id\hat{y}_i$. 

The two dimensional radial
amplitude of eq.(\ref{twod}) is now expanded in terms of the
CHH basis
\be
  \Phi_\alpha(x_i,y_i)=\rho^{\ell_\alpha+L_\alpha}
  f(r_{12})f(r_{23})f(r_{31}) \left[ \sum_k u^\alpha_k(\rho)\;\; 
  ^{(2)}P^{\ell_\alpha,L_\alpha}_k(\phi_i)\right],
\label{expk}
\ee
where the hyperspherical polynomials are given by ~\cite{rip1}
\be
^{(2)}P^{\ell_\alpha,L_\alpha}_k(\phi_i) = 
{\mathcal{N}}^{\ell_\alpha,L_\alpha}_k
\left(\cos{\phi_i}\right)^{\ell_\alpha}
\left(\sin{\phi_i}\right)^{L_\alpha} 
P^{l_\alpha+\frac{1}{2},L_\alpha+ \frac{1}{2}}_k(\cos{2\phi_i})
\label{polh}
\ee
with ${\mathcal{N}}^{\ell_\alpha,L_\alpha}_k$ a normalization
factor and $P_k^{a,b}$ a Jacobi polynomial. The quantum number
$k$ is a non negative integer related to the grand orbital quantum
number $K=\ell_\alpha+L_\alpha+2k$. The product of the
hyperspherical polynomial defined in eq.(\ref{polh}) times the
spherical harmonics coupled to $LM$ in eq.(\ref{twod}) gives a
standard three--body hyperspherical harmonic (HH) function with
defined total angular momentum.

The other ingredient in the expansion of eq.(\ref{twod}) is the
correlation factor, taken in the present work of the product (Jastrow)
type. Its role is to speed the convergence of the expansion describing
those configurations in which two particles are close to each other.
The use of Jastrow correlation factors has a long tradition in
the description of infinite systems as nuclear matter or 
liquid helium~\cite{fantoni} as well as in the description of
light nuclei ~\cite{chh}. The wave function describing
strongly interacting structures, in which the interaction is
highly repulsive at short distances, is practically zero when the
distance between any pair of particles is smaller than the repulsive core
of the potential. The correlation factor imposes this behaviour
as it can be seen from the specific form of the correlation function
$f(r)$ given in Fig.1. The short range behaviour of $f(r)$ is 
governed by the two-body potential whereas its medium and long
range form is not critical since the structure of the system will
be constructed by the HH basis. A simple procedure to determine
the correlation function for states in which the pair $(i,j)$ is
in a relative state with zero angular momentum is to solve the 
following zero-energy Schr\"odinger-like equation ~\cite{cfun}
\be
 [-{\hbar^2 \over m}({\partial^2\over \partial r^2}+ 
 {2 \over r}{\partial\over \partial r})+V(r)+ W(r)]f(r)=0,
\ee
where 
$V(r)$ is the He-He interaction used in the Hamiltonian of eq.(\ref{ham}).
The additional term  $W(r)$ is included to allow the
function $f(r)$ to satisfy an appropriate healing condition. It is
chosen as
\be
 W(r)=W_0\exp(-r/\gamma).
\ee
The specific value of $\gamma$ is not important provided that
the ranges of the additional potential $W(r)$ and $V(r)$
are comparable~\cite{cfun}. 
Hereafter its value has been fixed to $\gamma= 5$ \au. 
The depth $W_0$ is fixed requiring that
$f(r)\rightarrow 1$ for values of $r$ greater than the range of
the potential $V(r)$.

The hyperradial functions $u^\alpha_k(\rho)$ of
eq.(\ref{expk}) are taken as a product of a linear combination of 
Laguerre polynomials and an exponential tail:
\be
\label{eq:lexp}
u^\alpha_k(\rho)=\sum_m A^\alpha_{k,m}L^{(5)}_m(z)\exp(-{z\over 2})
\ee
where $z=\beta\rho$ and $\beta$ is a non-linear variational parameter.
Let $|\alpha,k,m>$ be a correlated completely symmetric element of the
expansion basis, where $\alpha$ denotes the angular channels and
$k,m$ are the indices of the hyperspherical and Laguerre polynomials,
respectively. In terms of the basis elements the wave function
(\ref{ampl}) results
\be
\Psi = \sum_{\alpha,k,m} A^\alpha_{k,m}|\alpha,k,m>.
\label{expf}
\ee
The problem is to determine the linear coefficients $A^\alpha_{k,m}$. The
wave function and energy of the different bound states are obtained
by solving the following generalized eigenvalue problem
\be
\sum_{\alpha',k',m'}A^{\alpha'}_{k',m'}<\alpha',k',m'|H-E|\alpha,k,m>=0.
\label{gep}
\ee
The dimension $N$ of the involved matrices is related to three indices:
the number of angular channels $N_\alpha$, the number of hyperspherical
polynomials per channel $K_\alpha$ and the number of Laguerre
polynomials per channel $M_\alpha$. According to the 
Hylleraas-Undheim-MacDonald's theorem  ~\cite{hyl,mcd}, there exists a 
one-to-one correspondence between the approximate energy levels $E_{i}(N)$ 
and the exact levels $\epsilon_{i} \equiv E_{i}(\infty) $, the $i$-th 
approximate level being an upper bound to the $i$-th exact level. 
Mathematically the following relations hold: \\
\be
E_{i+1}(N+1) \ge E_{i}(N) \ge E_{i}(N+1) 
\ee
and
\be
\lim_{N\rightarrow +\infty} E_i(N)=\epsilon_i
\ee
The implementation of the method in the specific case of the helium
trimer in which two bound states are known to exist,
consists in solving the generalized eigenvalues problem for 
increasing values of $N$, until a convergence is 
achieved in the estimates of the ground state $E_{0}$ and 
excited state $E_{1}$. Moreover,
an optimum choice of the non-liner parameter $\beta$ can be used 
to improve the pattern of convergence.

\section{Bound state calculations}

The generalized eigenvalue problem of eq.(\ref{gep}) can be solved
to find bound states of general value of total angular momentum $LM$. 
Here we are interested in the ground and excited state of the helium 
trimer both having total angular momentum $L=0$. In such a case
the angular dependence of each $i$-amplitude of the wave function 
reduces to a Legendre polynomial $P_l(\mu_i)$ with
$\mu_i={\hat x}_i\cdot{\hat y}_i$. Moreover, the angular
channel with $\ell_\alpha=L_\alpha=0$ is, by far, the most important
and it has been the first one to be considered. 
Contributions from successive channels, with $\ell_\alpha=L_\alpha>0$, 
are highly suppressed due to centrifugal barrier considerations
and can be safely disregarded as it will be discussed latter.

The matrix elements defined in eq.(\ref{gep}) have been obtained
numerically. In general,
as the dimension of the matrices increases, numerical problems could 
arise from integrals containing polynomials of high degree. In fact,
a high number of basis functions is expected in order to describe
simultaneously both the ground and excited state, which have a completely 
different spatial extent. On the other hand, the correlation
functions introduce a complicated structure at short distances.
Therefore, a dense grid of integration points is necessary.
The integrals have been performed in the set of coordinates
$[\rho,\phi_3,\mu_3]$ using a Gauss formula in the
variable $\mu_3$ and a Chebyshev Lobatto
formula in the variable $\cos(2\phi_3)$. Grids of 300 points for the
first case and 3000 for the second have been used.
In the variable $\rho$ the integrals have been performed on a scaled grid:
\bea
\left\{ \begin{array}{ll}
                \rho_0=h &  \\
                \rho_n= \chi\, \rho_{n-1} & (n=1,n_{max})
                 \end{array} \right. \eea
with the choice $h=.07$ \au, $\chi=1.008$, and $n_{max}\approx 800$, 
covering the range $0-5000$ \au. A numerical accuracy of $10^{-3}$ mK
has been obtained in the calculation of the binding energies.

The convergence of the eigenvalues has been studied increasing
the number of basis elements, restricting the discussion to one channel,
namely the $\ell_\alpha=L_\alpha=0$ channel. In this case
a totally symmetric wave function can be constructed for
values of the quantum number $k=0,2,....,k_{max}$
(no symmetric function exists for $k=1$). Therefore, the number
of hyperspherical polynomials $K_0$ included in a specific calculation
is $k_{max}$, except for $k_{max}=0$ which corresponds $K_0=1$. The number of
Laguerre polynomials is $M_0=m_{max}+1$ with 
$m_{max}$ the maximum degree considered. The
total dimension of the problem to be solved is $N=K_0 \cdot M_0$.

In Table II the convergence of $E_{0}$ and $E_{1}$ is shown
as a function of $k_{max},m_{max}$ for the LM2M2 potential. We observe that, 
while the ground state energy $E_0$ converges
with a rather small basis set, for the excited state $E_1$
it is needed a much bigger basis (about one order of magnitude bigger). 
The ground state converged
to the value $E_0=-126.36$ mK with $k_{max}=20$ and $m_{max}=20$
whereas the excited state converged to $E_1=-2.27$ mK with 
$k_{max}=80$ and $m_{max}=32$. In order to speed the convergence
with respect to the Laguerre polynomials the value of the non linear parameter
$\beta$ has been optimized. For the ground and excited state we have used 
$\beta=0.40$ \au$^{-1}$ and $\beta=0.10$ \au$^{-1}$ respectively.
 
After the convergence of the first channel is achieved, the
contribution of the channel $\ell_\alpha=L_\alpha=2$ can be evaluated.
The first four linearly independent totally symmetric basis elements belonging
to the second channel correspond to values of the grand angular quantum
number $K=12,16,18,20$. The inclusion of these
elements gives extremely small contributions to the binding energy of the
two states and does not change the estimates given above.
It is important to notice that the Jastrow correlation factor 
introduced contributions from channels with $\ell_\alpha=L_\alpha \ge 2$ 
already in a calculation limited to the first channel.

Special attention has been given to the study of the convergence with
the non-linear parameter $\beta$. In Fig.2 
we reported the ground and excited state energy curves as a function of 
$\beta$ for increasing values of $k_{max}$. The number of Laguerre 
polynomials has been kept fixed at $m_{max}=24$. For the ground state energy
the upper curve corresponds to $k_{max}=0$, i.e. only one hyperspherical
polynomial has been taken into account, and the lower curve corresponds to
$k_{max}=20$. Results with larger values of $k_{max}$ are not shown
since they completely overlap with the result at $k_{max}=20$.
For the excited state energy the different curves correspond to
$k_{max}=20,40,60,80$. 
We observed that there is a region where the variation of $\beta$
does not appreciably affect the binding energies. 

The variational method provides, in addition to an upper bound to the
exact energy of the states, a variational estimate of the
corresponding wave functions. Through the wave function it is
possible to calculate a certain number of mean values characterizing
the ground and excited state. In Tables III-IV,
we computed the binding energy, the mean value 
of the kinetic energy, the potential energy, the interparticle distance 
$r_{ij}$, and the distance $r_i$ between the $i$-particle and the
 center of mass. 
 The HFDB, LM2M2, TTY, SAPT1 and SAPT2 interactions have been
considered. Other than typical 
observables, we also computed the asymptotic normalization constants 
$c_\nu$ for the two bound states, 
as defined in Refs.~\cite{ca1,ca2} and briefly described below, 
and the percentage $P^\nu_d$ of dimer-like 
structures in the trimer wave functions $\Psi_\nu$ ($\nu=0,1$). 

The variational wave function is constructed in the present work
as a sum of correlated products of polynomials with an exponential
tail in the hyperradius. The short range behaviour of the wave function
is governed by the correlation factor whereas in the medium and asymptotic
region the expansion should reproduce the structure of the system.
It is interesting to evaluate the flexibility of the correlated basis
to reproduce correctly the asymptotic behaviour.
Let us introduce the asymptotic function $\Theta_\nu(y)$:
\be
\Theta_\nu(y) = \sqrt{2 q_\nu}\frac{e^{-q_\nu y}}{y} 
\ee
where $q_\nu=\sqrt{m \, |E_\nu-E_d|/ \hbar^2}$, and $E_\nu$ and $E_d$ are 
respectively the trimer ($\nu=0,1$) and the dimer binding energies. Here  
$y$ denotes the second Jacobi coordinate defined in eq.(\ref{jacs}).
 In the configuration in which one atom is far from the other
 two the trimer w.f. for the
ground and excited state $\Psi_\nu$ behaves asymptotically like:
\be
\label{eq:asy1}
\Psi_\nu \stackrel{y\rightarrow\infty}{\longrightarrow} 
     c_\nu \Theta_\nu(y) Y_{00}({\hat y})\Phi_d(\xv)
\ee
where $\Phi_d$ is the dimer w.f., and $c_\nu$ the asymptotic constant.
Therefore, the overlap function 
\be
{\cal O}_\nu(y)=\int\Phi^*_d(\xv)Y^*_{00}({\hat y})\Psi_\nu d{\hat y}d\xv
\ee
is proportional to $e^{-q_\nu y}/{y}$ as $y\rightarrow\infty$. In Fig.3
we plot the overlap functions ${\cal O}_\nu(y)$ and the asymptotic functions
$\Theta_\nu(y)$ for both the ground and excited state. From the figure
it is clear that the two curves approach to each other as the
distance $y$ increases. We also observe the very large extension of the
excited state. The asymptotic constants $c_\nu$ are obtained evaluating
the ratio ${\cal O}_\nu(y)/\Theta_\nu(y)$ at large $y$ values and are
given in Tables III and IV for the different interactions considered.
For the ground state the five interactions produce similar values
of $c_0$, though the values of LM2M2 and TTY are slightly smaller.
Conversely, for the excited state the result obtained with SAPT2
is smaller than that obtained with the other potentials.

The percentage $P_d$ of a dimer-like structure inside the trimer
is defined as:
\be
P^\nu_d=\int_0^\infty \left| {\cal O}_\nu(y) \right|^2 y^2 dy \ .
\ee
The results for the ground and excited state are collected in Tables III 
and IV for the different interactions. For the trimer ground state the
probability of a dimer-like structure is around $0.35$, whereas for
the excited state this probability increases up to $0.75$ for
LM2L2 and TTY interactions. The two SAPT interactions predict a lower
dimer-like structure, in particular SAPT2. This behaviour is related
to the slightly tighter binding predicted by the SAPT potentials for
the two bound states. As a general remark, the
very high value of $P^1_d$ (nearly $70\%$, compared to $35\%$ 
of the ground state) suggests that the excited state of \trimer can 
effectively be pictured as a third particle orbiting around a two particle 
structure. 

In Fig.4 we plotted some distribution functions relevant to 
understand the structure of the two bound states of the trimer. 
Namely, we plotted the pair distribution function $p(r_{ij})$, 
which represents the probability to find the particles $i$ and $j$ at 
distance $r_{ij}$, and the mass distribution function $m(r_i)$, which 
is related to the probability to find the particle $i$ at distance $r_i$ 
from the center of mass of the system.

Our results for the ground state agree quite well with the ones published 
in literature \cite{esry,gnt2,nielsen,esry2,gnt1}. For the SAPT potentials we 
find that the bond becomes slightly tighter ($\approx$ 5\%), as a result
of the its more attractive well. 
But basically there are not qualitative differences in 
describing the trimer with any of the different pairwise interactions. 
There is a discrepancy in literature whether the main spatial arrangement of 
the three particles in the ground state is either a quasilinear or 
equilateral configuration \cite{comm}. Our results seem to agree with
 the latter. In fact, we can try to discriminate between the two by looking
 at the pair distribution and the mass distribution functions. 
As it is shown in Fig.4, the probability to find any particle 
in proximity of the CM is almost zero. This strongly suggests that the most 
probable configuration is the equilateral one. Moreover, the ratio between 
$<r_{ij}>/<r_i>$ is very close to the ideal $\sqrt{3}$ of equilateral 
triangle for all the potentials we used. Regarding the excited state, the 
presence of a two peaks structure in the pair distribution function agrees
with the interpretation of such a state as composed by a two particle core
 surrounded by the third atom at a larger distance.

For the excited state the results do not depend qualitatively on the 
potential we use. This may look very surprising at a first glance, 
because this state is suspected to be an Efimov state, and consequently 
it is expected to be strongly affected by any minimum variation of the 
pairwise interaction.
Following Refs.\cite{esry,gnt2,gnt1}, we studied the 
behaviour of \trimerecc as a function of the strength of the pairwise 
interaction. In Fig.5 we plotted the energy difference $E_1-E_d$ of 
the system as a function of the parameter $\lambda$ defined by:
\be
V_{He-He}= \lambda V_L
\ee
where $V_L$ is the LM2M2 potential. We found that this state disappears
both increasing and decreasing $\lambda$, in agreement with the claim
that it is an Efimov state. Our results agree quite well with
Refs.~\cite{esry,gnt2}, where the peculiarity of such a behaviour
has been widely discussed. To summarize it, we observed that the
trimer begins to support an excited state at $\lambda \approx 0.975$;
then, increasing $\lambda$, the binding energy firstly increases,
until it achieves its maximum at $\lambda \approx 1.05$, and successively 
decreases, until it dissociates.

In order to compare the different pairwise interactions we assume that
due to the very  large extension of the w.f. compared to the range of the 
potential, the particles are not sensitive to the particular shape of it, 
but somehow to its average strength. Accordingly we define

\be
\lambda_x= {\int^\infty_{\sigma_x} V_x \over \int^\infty_{\sigma_L} V_L} \ ,
\ee
with $x=L,T,H,S1,S2$ in accordance with the LM2M2, TTY, HFDB, SAPT1 and
SAPT2 interactions and $\sigma_x$ is the interparticle distance where 
the considered potential changes sign, i.e. $V_x(\sigma_x)=0$.
In the smaller frame of Fig.5 we reported the different values of 
$\lambda_x$. It is worth to observe 
that the potentials do not differ so much to show dissimilar 
results, as all the points lie in a small interval of $\lambda_L=1$. In
fact a plot in function of $\lambda$ for the other interactions shows that
in all cases the physical case $\lambda=1$ is on the left of the minimum
of the curve, as for the LM2M2 potential.

The most peculiar feature of an Efimov like system is that it disappears 
tightening the interaction among its components. Physically, such a behaviour 
could be explained by picturing the system like composed by a third particle 
orbiting around a two particle sub-system. Increasing the strength of the 
pairwise interaction makes the two particles tighter to each other, and the 
third one evaporates as a result of its very weak bound. 

\section{conclusions}

In the present work the helium trimer has been investigated using
the most recent helium-helium potential models.
The helium trimer wave function has been
expanded in terms of the CHH basis. Then, the energies and wave functions
of the ground and excited state have been obtained
by solving a generalized eigenvalue problem. The Hylleraas-Undheim-MacDonald's
theorem assures that the obtained results for the energy of the levels
represent upper bounds to the exact values. 

The strong repulsion of the He-He potential at short distances engenders
some difficulties in the description of the three atoms system in terms of an
expansion basis. Very large bases are then necessary in order to obtain a
 satisfactory description of the structure of both bound states. 
The structure is such that the probability
of finding two atoms at short distances is close to zero and this type
of behaviour is difficult to describe using, for example, a polynomial
expansion. Correlation factors naturally introduce this behaviour
accelerating the rate of convergence of the expansion
basis. In particular the CHH basis combines a Jastrow correlation
factor with the HH basis. The CHH basis has been used before in the
description of nuclear systems~\cite{chh} in order to take into account the
strong repulsion of the nucleon-nucleon interaction at short distances.
Here, the CHH basis has been used to study five different
interactions in the description of the trimer. The pattern of
convergence for the bound and excited state has been studied by
increasing the number of basis elements. With a sufficient number of
elements, the dependence on the non linear parameter $\beta$ is
smooth. Therefore it is possible to obtain a simultaneous description
of the bound states with high accuracy.
The results are collected in Tables III-IV and are in close
agreement with previous results obtained by different groups using the
HFDB, LM2M2 and TTY interactions. The estimates for the binding energy
are upper bounds to the exact levels and show that the
variational method can be used to describe strongly correlated systems, 
as helium trimer, with results that are believed to be among the most accurate
ones at present.

Some interesting aspects of the wave function have been studied. Its
asymptotic behaviour in a configuration where one atom is moving away 
from the other two
is given in eq.(\ref{eq:asy1}). In Fig.3 this behaviour is shown
for the two bound states. From this study the asymptotic constants $c_\nu$
have been extracted. In some particular systems the asymptotic constants
can be measured~\cite{ludwig}. Moreover, the probability $P_d$ of a
dimer-like structure inside the trimer has been calculated. This
quantity gives a clear idea of the spatial structure of the molecule. For
the ground state we obtained $P_d^0 \approx 0.35$ whereas for the excited
state $P_d^1 \approx 0.70$. This latter result suggests a configuration
of two atoms in a dimer-like state with a third atom orbiting.

The present work should be consider as a first step in the use of
the variational technique with correlated basis functions for describing
small helium clusters. The extension of the method to study larger systems 
is feasible. The study of the bound states of the tetramer is at present
underway and will be the subject of a forthcoming paper.

\begin{acknowledgements}
The authors would like to acknowledge Prof. L. Bruch for helpful discussions.
\end{acknowledgements}


\newpage 

\begin{table}[h]
\begin{center}
\begin{tabular}{cccccccc} 
  & &HFDB&LM2M2&TTY&SAPT1&SAPT2& Ref.  ~\cite{gris} \\ \hline
 $\epsilon$ & (K) & $10.95$  & $10.97$ & $10.98$ &  $11.05$ & $11.06$ & \\ 
 $r_m$ & (\au) & $5.599$ & $5.611 $ & $5.616$  & $5.603$ & $5.602$ & \\
$\sigma$ & (\au) & $4.983$ & $4.992$ &$5.000$  & $4.987$ & $4.987$ & \\ \hline
 $|E_b|$ & (mK)  & $1.685$ & $1.303$ & $1.313$  & $1.733$ & $1.816$ & \\ 
$\bra T \ket$ & (mK) & $112.2$ & $99.43$ & $99.93$ & $115.0$ & $117.8$ & \\ 
$-\bra V \ket$ & (mK) & $113.9$ & $100.7$ & $101.2$ & $116.8$ & $119.6$ & \\ 
 $\bra R \ket $ & (\au) & $87.81$ & $97.96$ & $97.62$ & $86.04$ & $84.24$ & $98\pm 8$\\ 
  $\sqrt{\bra R^2 \ket}$ & (\au) & $119.0$ & $132.9$ & $132.5$ & $116.5$ & $114.0$ & \\
 a & (\au)& $170.5$ & $191.4$ & $190.7$ & $166.9$ & $163.2$ &  \\ \hline 
 $\hbar^2/4m\bra R \ket ^2$ & (mK)  & $1.403$ & $1.127$ & $1.136$ & $1.462$ & $1.525$ & $1.1+0.3/-0.2$\\ 
 $2\bra R \ket$&(\au)&$175.6$&$195.9$ & $195.2$ & $172.1$ & $168.5$ & $197+15/-34$ \\ 
\end{tabular}
\caption{Characteristic values of the different potentials and their 
relative bound states. $R$ represents the He-He distance,  $\epsilon$ is the 
strength of the potential at its point of  minimum $r_m$ and $\sigma$ is the 
distance at which the potential changes sign.}
\end{center}
\end{table}

 \begin{center}
 \begin{table}[h]
 \begin{tabular}{|c|ccccc|} 
  & \multicolumn{5}{c|}{ground state} \\ 
\hline
 $k_{max}$ & 5 & 10 & 15 & \multicolumn{2}{c|}{20}  \\ 
 $m_{max}$ &$E_0$(mK)&$E_0$(mK)&$E_0$(mK)&$E_0$(mK)&$T_0$(mK)  \\
\hline
  4  &-120.376&-120.689&-120.726&-120.737&1769.481 \\
  8  &-126.080&-126.274&-126.286&-126.288&1662.829 \\
  12 &-126.143&-126.337&-126.348&-126.349&1660.232 \\
  16 &-126.148&-126.342&-126.353&-126.354&1660.185 \\
  20 &-126.149&-126.343&-126.354&-126.355&1660.186 \\
  24 &-126.149&-126.343&-126.354&-126.355&1660.187 \\
\hline
  & \multicolumn{5}{c|}{excited state} \\ 
\hline
 $k_{max}$ & 20 & 40 & 60 & \multicolumn{2}{c|}{80}  \\ 
 $m_{max}$ &$E_1$(mK)&$E_1$(mK)&$E_1$(mK)&$E_1$(mK)&$T_1$(mK)  \\
\hline
  8         &-1.168&-1.579 &-1.612 &-1.622&139.535  \\
  12        &-1.523&-2.097 &-2.148 &-2.160&127.139  \\
  16        &-1.555&-2.150 &-2.222 &-2.237&123.526  \\
  20        &-1.562&-2.157 &-2.237 &-2.257&122.247  \\
  24        &-1.565&-2.160 &-2.240 &-2.262&121.943  \\
  28        &-1.567&-2.161 &-2.241 &-2.264&121.927  \\
  32        &-1.567&-2.162 &-2.242 &-2.265&121.935  \\
 \end{tabular}
 \caption{Convergence of the LM2M2 ground state energy $E_0$ and
 excited state energy $E_1$ for increasing values of the order of
 hyperspherical polynomials $k_{max}$ and Laguerre polynomials $m_{max}$.
 In the last column the convergence for the kinetic energy is shown.
 Basis states with $\ell_\alpha=L_\alpha=0$ have been considered. The
 non linear parameter $\beta$ has been fixed to $0.40$ \au$^{-1}$ for
 the ground state and $0.10$ \au$^{-1}$ for the excited state. 
 }
 \end{table}
 \end{center}

\begin{table}[h]
\begin{center}
\begin{tabular}{ccccccc} 
 & & HFDB & LM2M2 & TTY & SAPT1 & SAPT2 \\ \hline
 $B$ & (mK) & $133.0$ & $126.4$ & $126.4$   & $133.8$ & $135.1$ \\
 $<T>$ & (mK) & $1698$  & $1660$ & $1662$ & $1707$ & $1715$ \\ 
 $-<V>$ & (mK) & $1831$  & $1787$ & $1788$ &  $1841$ & $1850$ \\
 $<r_i>$ & (\au) & $10.38$ & $10.51$ & $10.49$  & $10.36$ & $10.25$ \\ 
 $\sqrt{<r_i^2>}$ & (\au) & $12.11$ & $12.28$ & $12.26$  & $12.09$ & $12.03$\\
 $<r_{ij}>$ & (\au) & $17.98$ & $18.21$ & $18.16$  & $17.95$ & $17.77$ \\ 
 $\sqrt{<r_{ij}^2>}$ & (\au) & $20.97$ & $20.71$ & $20.71$  & $20.95$ & 
 $20.84$ \\ 
 $c_0$ & & $1.22$ & $1.17$ & $ 1.18 $  & $ 1.24 $ & $1.25$ \\ 
 $P^0_d$ & & $0.3614$ & $0.3310$ & $ 0.3311 $  & $0.3619 $ & $0.3539$ \\ 
\end{tabular}
\caption{ Binding energy and mean values of the kinetic and potential
energy of the helium trimer ground state calculated for different
pairwise interactions. The mean values and square root mean values of
the distance $r_i$ of particle $i$ from the CM,
and the interparticle distance $r_{ij}$ are also given. In the last
two rows the asymptotic constant $c_0$ and the probability of a 
dimer-like structure are reported.}
\label{p1}
\end{center}
\end{table}

\begin{table}[h]
\begin{center}
\begin{tabular}{ccccccc} 
 & & HFDB & LM2M2 & TTY & SAPT1 & SAPT2 \\ \hline 
 $B$ & (mK) & $2.735$ & $2.265$ & $2.277$  & $2.788$ & $2.885$ \\ 
 $<T>$ & (mK) & $134.1$ & $121.9$ & $122.4$  & $135.7$ & $137.8$ \\ 
 $-<V>$ & (mK) & $136.8$ & $124.2$ & $124.7$  & $138.5$ & $141.7$ \\
 $<r_i>$ & (\au) & $87.24$ & $94.00$ & $93.67$  & $83.07$ & $76.22$ \\ 
 $\sqrt{<r_i^2>}$ & (\au) & $103.5$ & $111.1$ & $109.5$  & $96.67$ & $81.00$\\ 
 $<r_{ij}>$ & (\au) & $145.6$ & $157.0$ & $150.7$  & $139.1$ & $125.0$ \\ 
 $\sqrt{<r_{ij}^2>}$ & (\au) & $177.7$ & $192.5$ & $183.7$  & $167.5$ & 
 $141.1$ \\
 $c_1$ &  & $1.20$ & $1.24$ & $1.24$  & $1.19$ & $1.09$ \\ 
 $P^1_d$ &  & $0.7266$ & $0.7462$ & $0.7461$  & $0.7030$ & $0.5816$ \\ 
\end{tabular}
\label{p2}
\caption{ Binding energy and mean values of the kinetic and potential
energy of the helium trimer excited state calculated for different
pairwise interactions. The mean values and square root mean values of
the distance $r_i$ of particle $i$ from the CM,
and the interparticle distance $r_{ij}$ are also given. In the last
two rows the asymptotic constant $c_0$ and the probability of a 
dimer-like structure are reported.}
\end{center}
\end{table}

\noindent Figure Captions

\noindent Figure 1. 
The LM2M2 potential $V$, its corresponding
ground state wave function $\Phi_d$ as well as the correlation function $f$
as a function of the interparticle distance $r$.

\noindent Figure 2. 
The ground state energy $E_0$ and excited state energy $E_1$
as a function of the non linear parameter $\beta$. 
For the ground state the curves corresponding to $k_{max}=0,20$ are
shown, whereas for the excited state the curves are
given in correspondence to
$k_{max}=20,40,60,80$. The number of Laguerre polynomials has been fixed
to $m_{max}=24$ in all cases.

\noindent Figure 3.
Overlap functions ${\cal O}_\nu(y)$ (solid line) and asymptotic
functions $\Theta_\nu(y)$ (dashed line) for the ground state 
($\nu=0$) and excited state ($\nu=1$) .

\noindent Figure 4. 
Distribution functions $p(r_{ij})$ and $m(r_i)$ for the ground
and excited state of the helium trimer.

\noindent Figure 5. 
The energy difference $E_1-E_d$ as a function of $\lambda$. 
In the small frame the positions in the curve of the different values
for $\lambda_x$ calculated as explained in the text are given.


\begin{thebibliography}{9}

\bibitem{luo2} F. Luo, C. F. Giese and W. R. Gentry, 
               J. Chem. Phys. {\bf 104}, 1151 (1996)
\bibitem{schol} W. Schoellkopf and J. P. Toennies, 
               J. Chem. Phys. {\bf 104}, 1155 (1995)
\bibitem{schol2} W. Schollkopf and J. P. Toennies, 
                 science {\bf 266}, 1345 (1994)
\bibitem{gris} R. Grisenti, W. Schoellkopf, J. P. Toennies,
               G. C. Hegerfeldt, T. Kohler and M. Stoll, 
               Phys. Rev. Lett. {\bf 85}, 2284 (2000)
\bibitem{aziz5} A. R. Janzen and R. A. Aziz, 
                J. Chem. Phys. {\bf 103}, 9626 (1995)
\bibitem{aziz3} R. A. Aziz, F. R. W. McCourt and C. C. K. Wong, 
                Molecular Physics {\bf 61}, 1487 (1987)
\bibitem{aziz1} R. A. Aziz and M. J. Slaman, 
                J. Chem. Phys. {\bf 94}, 8047 (1991)
\bibitem{tang1} K. T. Tang, J. P. Toennies and C. L. Yiu, 
                Phys. Rev. Lett. {\bf 74}, 1956 (1995)
\bibitem{aziz4} A. R. Janzen and R. A. Aziz, 
                J. Chem. Phys. {\bf 107}, 914 (1997)
\bibitem{kor1} T. Korona, H. L. Williams, R. Bukowski, B. Jeziorski and 
               K. Szalewicz, 
               J. Chem. Phys. {\bf 106}, 5109 (1997)
\bibitem{ef2} V. Efimov, Phys. Lett. {\bf 33B}, 563 (1970)
\bibitem{ef1} V. Efimov, Sov. J. Nucl. Phys. {\bf 12}, 589 (1971)
\bibitem{lew} M. Lewerenz, J. Chem. Phys.  {\bf 106}, 4596 (1997)
\bibitem{sof} A. K. Motovilov, W. Sandhas, S. A. Sofianos, E. A. Kolganova, 
              Eur. Phys. J. D {\bf 13}, 33 (2001)
\bibitem{bruch} L. W. Bruch, J. Chem. Phys. {\bf 110}, 2410 (1999);
           T.G. Lee, B.D.Esry, B.C. Gou, and C.D. Lin, J. Phys. {\bf B34},
            L203 (2001) 
\bibitem{esry} B. D. Esry, C. D. Lin and Chris H. Greene, 
               Phys. Rev. A {\bf 54} 394 (1996) 
\bibitem{gnt2} T. Gonz\'alez-Lezana, J. Rubayo-Soneira, S.Miret-Art\'es,
               F. A. Gianturco, G. Delgado-Barrio and P. Villareal, 
               Phys. Rev. Lett. {\bf 82}, 1648 (1999)
%
\bibitem{chh} A. Kievsky, M. Viviani and S. Rosati,
              Nucl. Phys. {\bf A551}, 241 (1993); 
              M. Viviani, A. Kievsky and S. Rosati, 
              Few-Body Syst. {\bf 18}, 25 (1995)
\bibitem{rip1} M. Fabre de la Ripelle, Annals of physics {\bf 123}, 185 (1979)
\bibitem{fantoni} S. Fantoni and A. Fabrocini, in {\sl Microscopic
 Quantum Many-Body Theories and Their Applications}, ed. by J. Navarro and
 A. Pols, Spingre-Verlag 1998, p. 119.
\bibitem{cfun} S. Rosati, M. Viviani and A. Kievsky, 
               Few-Body Syst. {\bf 9}, 1 (1990); 
               M. Viviani, A. Kievsky and S. Rosati, 
               Nuovo Cim. {\bf 105A}, 1473 (1992)
\bibitem{hyl} E. A. Hylleraas and B. Undheim, Z. Phys {\bf 65}, 759 (1930)
\bibitem{mcd} J. K. L. MacDonald, Phys. Rev. {\bf 43}, 830 (1933)
\bibitem{ca1} J. L. Friar, B. F. Gibson, D. R. Lehman and G. L. Payne, 
              Phys. Rev. C {\bf 25}, 1616 (1982)
\bibitem{ca2} H. Kameyama, M. Kamimura and Y. Fukushima, 
              Phys. Rev. C {\bf 40}, 974 (1989)
\bibitem{nielsen} E. Nielsen, D. V. Fedorov and A. S. Jensen, 
                  J. Phys. B {\bf 31}, 4085 (1998)
\bibitem{gnt1} T. Gonz\'alez-Lezana, J. Rubayo-Soneira, S.Miret-Art\'es,
               F. A. Gianturco, G. Delgado-Barrio and P. Villareal, 
               J. Chem. Phys. {\bf 110}, 9000 (1999)
\bibitem{esry2} D. Blume, Chris H. Greene and B. D. Esry, 
                J. Chem. Phys. {\bf 113} 2145 (2000) 
\bibitem{comm} B.D. Esry, C.D. Lin, C.H. Greene, and D. Blume, Phys. Rev. Lett.
        {\bf 86}, 4189 (2001); T. Gonz\'alez-Lezana {\sl et al.},Phys. Rev.
    Lett. {\bf 86}, 4190 (2001) 
\bibitem{ludwig} B. Kozlowska, Z. Ayer, R. K. Das, H. J. Karwowski and
                 E. J. Ludwig, Phys. Rev. {\bf C50}, 2695 (1994)

\end{thebibliography}
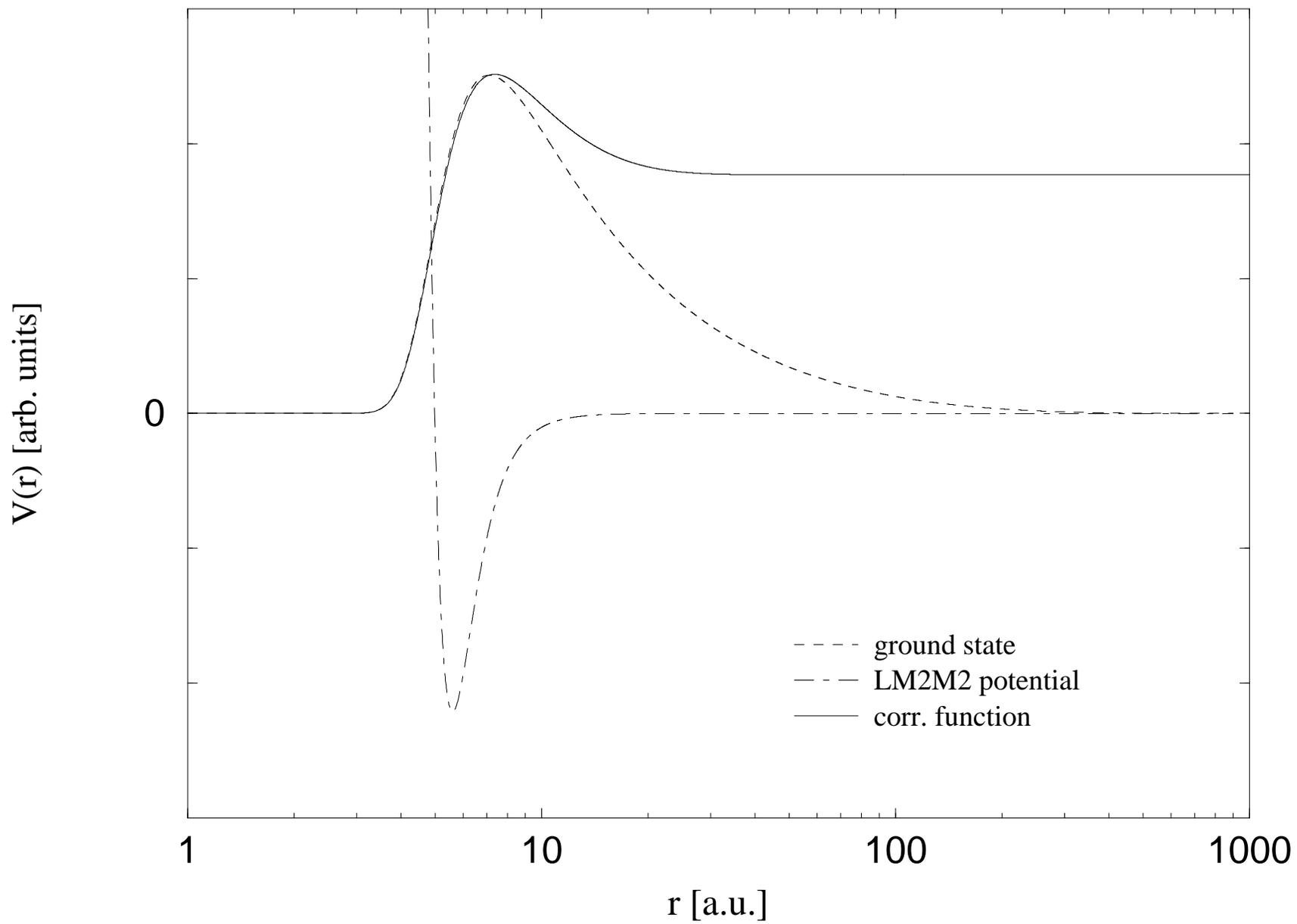
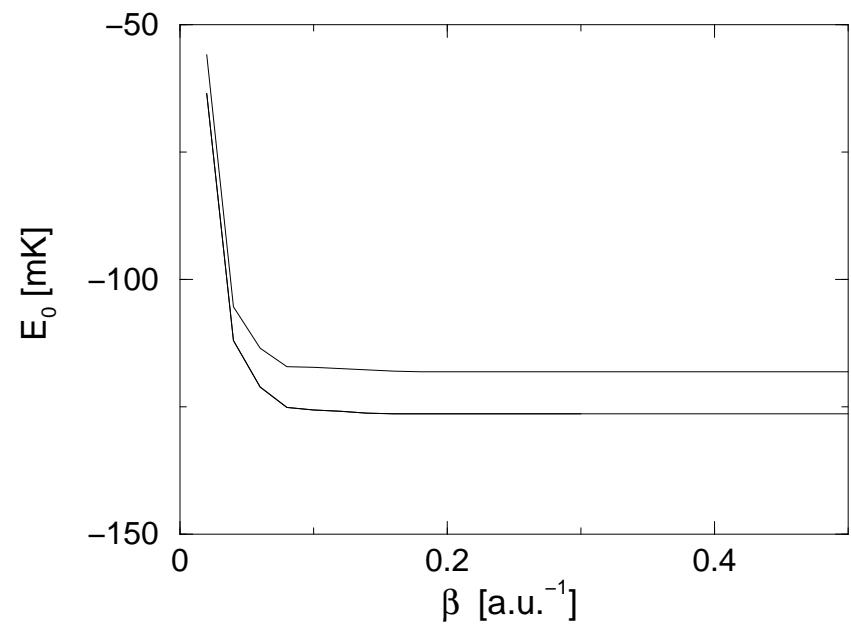
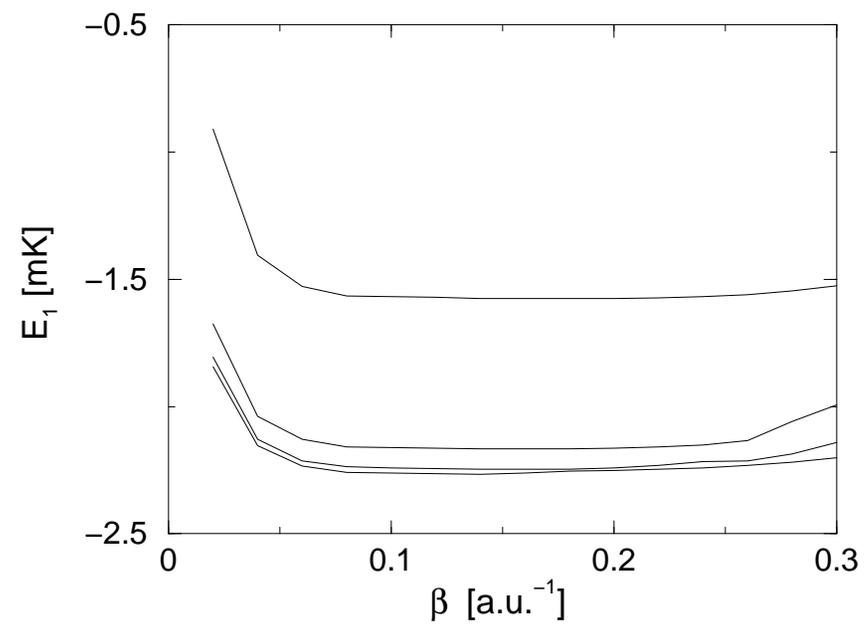
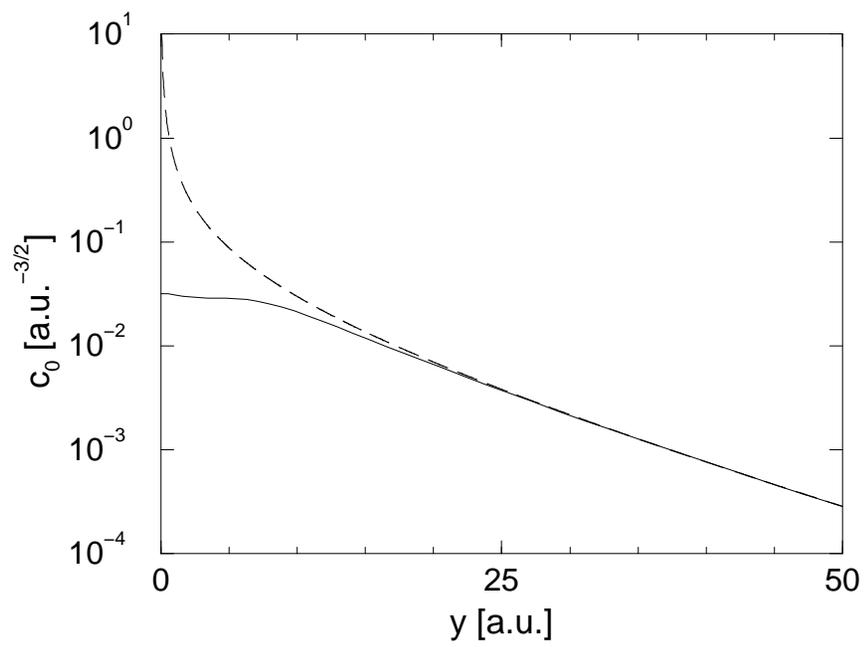
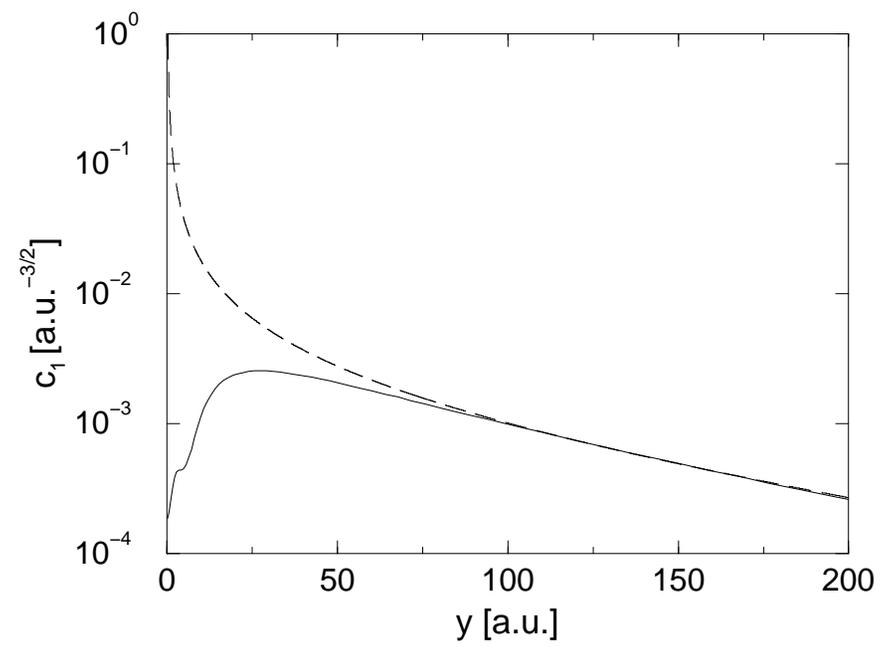
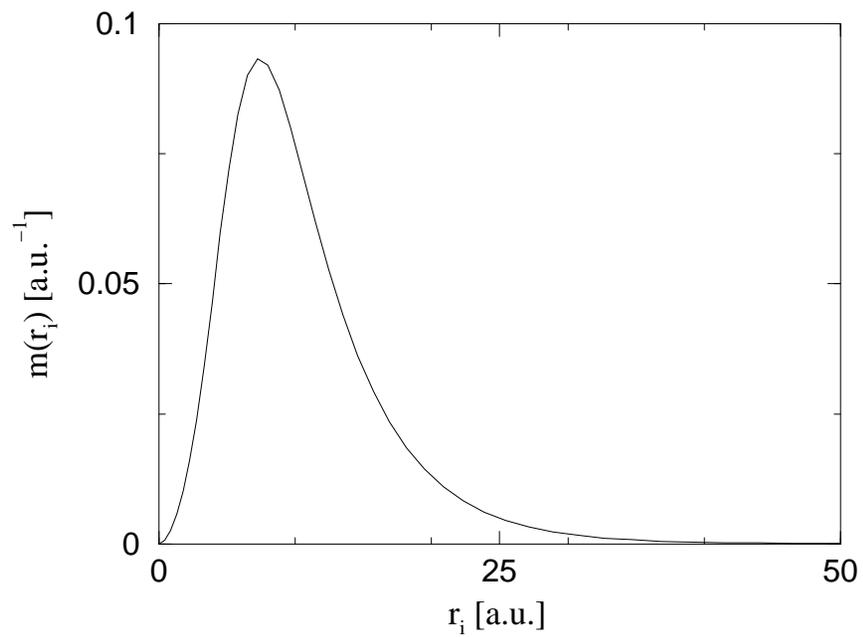
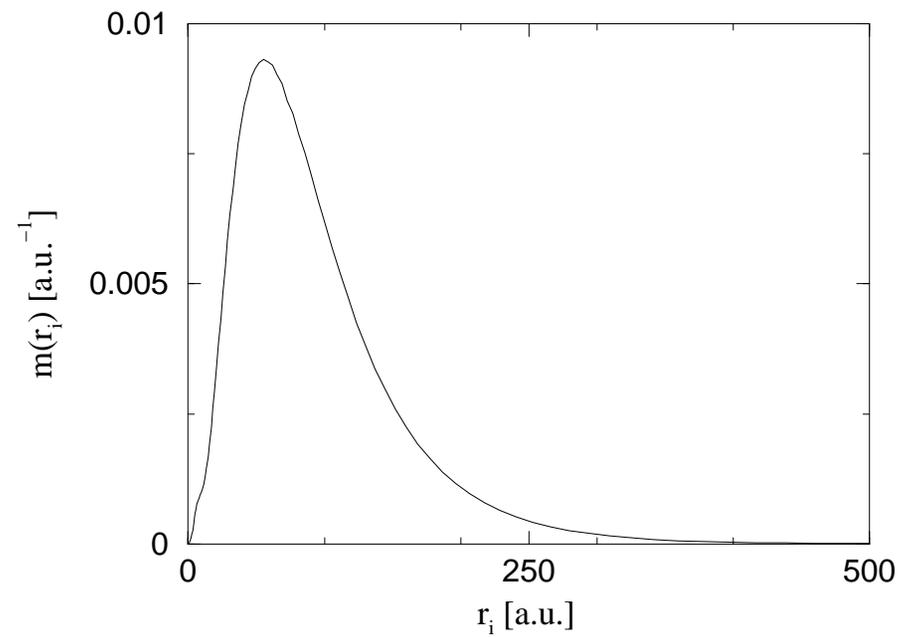
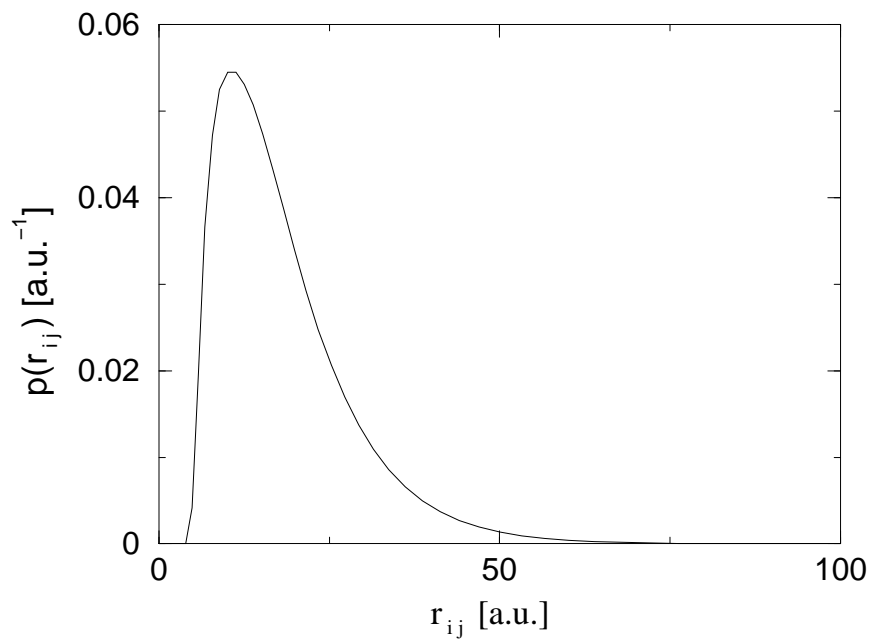
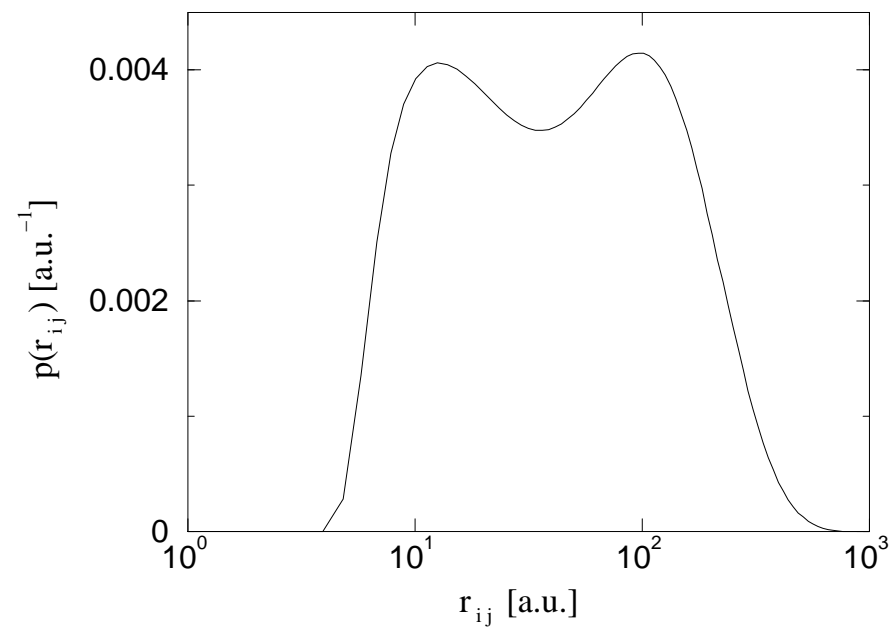
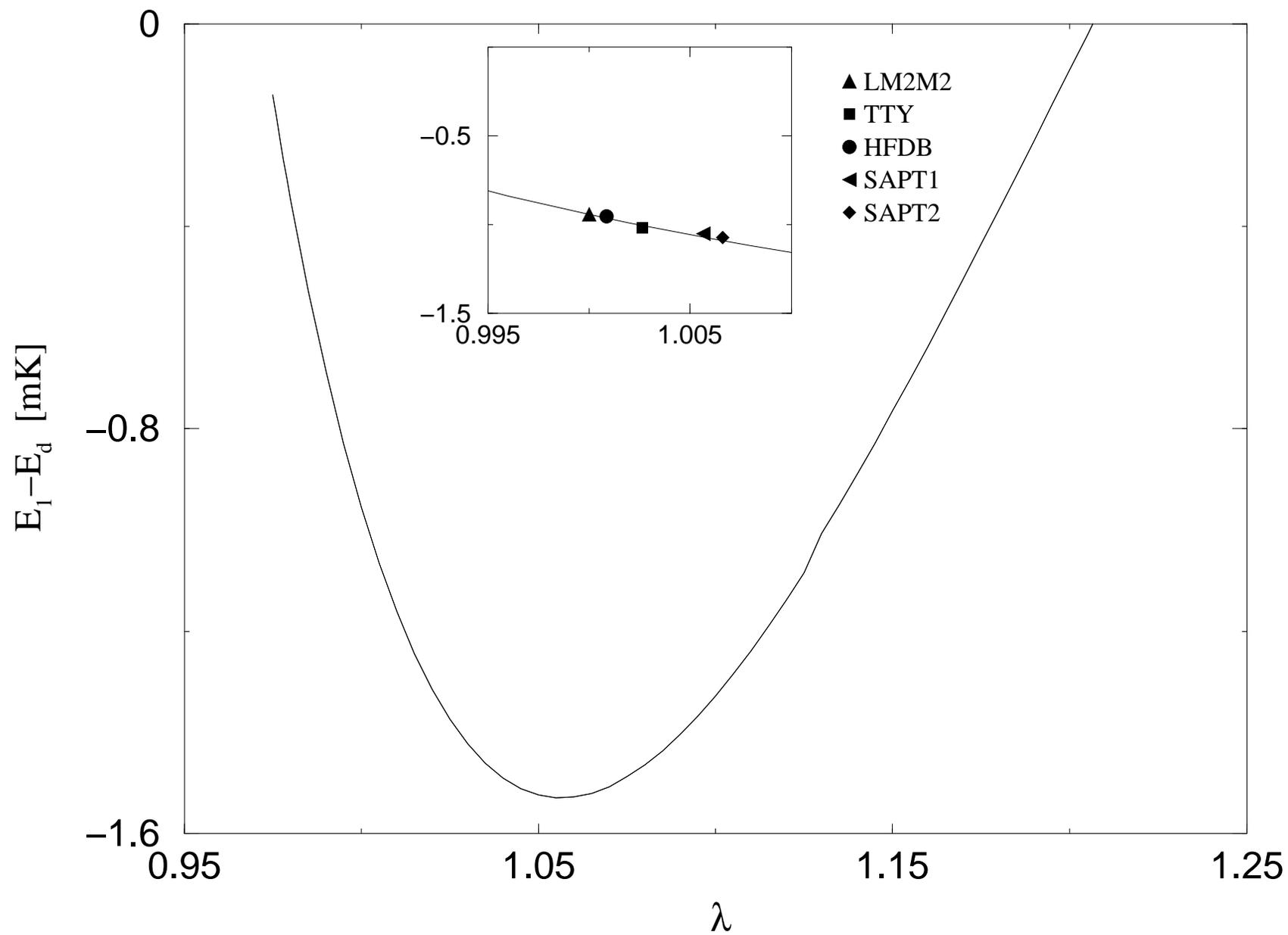
\end{document}